 \documentclass[final,3p,times,usenatbib]{elsarticle}

\usepackage[english]{babel}
\usepackage[latin1]{inputenc}
\usepackage{amsfonts} 
\usepackage{amsmath}
\usepackage{amssymb}
\usepackage{amsthm}
\usepackage{hyperref} 
\usepackage{graphicx} 
\usepackage[usenames,dvipsnames]{color}
\usepackage{rotating}
\usepackage{epsfig} 
\usepackage{graphicx}
\usepackage{threeparttable}
\usepackage{multirow}
\usepackage{url}
\usepackage{booktabs} 
\usepackage{lineno}

\usepackage{rotating}
\usepackage{pdfpages}
\usepackage{listings} 
\usepackage{multirow}
\usepackage{enumitem}

\definecolor{darkpastelgreen}{rgb}{0.01, 0.75, 0.24}
\definecolor{forestgreen}{rgb}{0.13, 0.55, 0.13} 
\definecolor{limegreen}{rgb}{0.2, 0.8, 0.2}
\definecolor{lavenderindigo}{rgb}{0.58, 0.34, 0.92}
\definecolor{magenta}{rgb}{1.0, 0.0, 1.0}
\definecolor{light-gray}{gray}{0.95}

\journal{Journal  }

\begin{document}

\begin{frontmatter}


\title{Design of phase III trials with long-term survival outcomes based on short-term binary results}



\author{Marta Bofill Roig\fnref{label1}}
\ead{marta.bofill.roig@upc.edu}
\author{Yu Shen\fnref{label2}}
\ead{yshen@mdanderson.org}
\author{Guadalupe G\'omez Melis\fnref{label1}}  
\ead{lupe.gomez@upc.edu}

\address[label1]{Departament d'Estad\'{i}stica i Investigaci\'{o} Operativa, Universitat Polit\`{e}cnica  de Catalunya, Barcelona, Spain}
\address[label2]{Department of Biostatistics, The University of Texas MD Anderson Cancer Center, Houston, Texas, U.S.A.\fnref{label3}}

\begin{abstract} 
Pathologic complete response (pCR) is a common primary endpoint for a phase II trial or even accelerated approval of neoadjuvant cancer therapy. 
If granted, a two-arm confirmatory trial is often required to demonstrate the efficacy with a time-to-event outcome such as overall survival. However, the design of a subsequent phase III trial based on prior information on the pCR effect is not straightforward. \\ 
Aiming at designing such phase III trials with overall survival as primary endpoint using pCR information from previous trials, we consider  a mixture model that incorporates both the survival and the binary endpoints. We propose to base the comparison between arms on the difference of the restricted mean survival times, and show how the effect size and sample size for overall
survival rely on the probability of the binary response and the survival distribution by response status,   both for each treatment arm.  
Moreover,  
we provide  the  sample size calculation under different scenarios and accompany them with the R package survmixer
where all the computations have been implemented. 
We  evaluate our proposal with a simulation study, and  illustrate its application through a neoadjuvant breast cancer trial. 
\end{abstract}


\end{frontmatter}




\section{Introduction} 

Neoadjuvant therapy, hoping to shrink a tumor before surgery, has become increasingly common for early-stage breast cancer. Examples of neoadjuvant therapy include chemotherapy, radiation therapy, and hormone therapy. Neoadjuvant therapy  permits breast conservation in patients who otherwise require mastectomy; enables direct evaluation of the tumor response, which may add prognostic information; and  allows for the examination of tissue, imaging, and biomarkers from biopsy.\cite{FDA2014}

In the context of early-stage breast cancer, the use of the binary endpoint pathologic complete response (pCR), defined as the complete eradication of invasive cancer,  has been proposed as an endpoint for accelerated approval by regulatory agencies. 
Trials under accelerated pathways are allowed to  
base  the benefit  on surrogate or intermediate endpoints. If granted, a confirmatory trial is still needed to demonstrate the efficacy based on  long-term endpoints, such as overall survival (OS) or event-free survival (EFS).\cite{FDA2014,EMA2014}

The association between pCR and survival endpoints has been extensively discussed in recent years.
A meta-analysis of existing randomized clinical trials and some cohort studies on neoadjuvant treatments showed that pCR in non-metastatic HER2-positive patients was associated with longer times to recurrence or death.\cite{DeMichele2015,Cortazar2014,Broglio2016} However, Cortazar et al\cite{Cortazar2014} insightfully noted that higher pCR rates due to an intervention  could not be used as a surrogate endpoint for improved EFS and OS at the trial-level analyses; indeed, little association was observed between an increased number of pCR responses and improved OS or EFS.  
Whereas at an individual patient level pCR endpoint could be strongly associated with long-term EFS and OS endpoints,\cite{Broglio2016} this is not yet a proven correlation at the trial level. How to predict  a beneficial treatment effect on survival endpoints based on  pCR improvement at trial level is nontrivial.

Several challenges arise when using pCR as a surrogate endpoint for a survival endpoint in the design of a phase III trial.  
Hatzis et al\cite{Hatzis2016} discuss how a large pCR treatment effect may not be translated into a similar long-term survival effect at the trial level. 
The authors pointed out that modest or even large improvements in pCR rate may only translate into small improvements in  survival endpoints, particularly in a patient population with good long-term prognoses.	They evaluate, through simulation,  the relationship between increased pCR rate and survival endpoints with different baseline prognoses under two critical assumptions: (i) the survival distributions of patients with pCR are the same, regardless which treatment was received; and (ii) the survival function of patients who do not achieve pCR is also similar in both arms. These two assumptions imply that the survival benefit from the new therapy arm is only derived from the improved pCR rate over the control arm. 
Whereas these assumptions could be reasonable for triple-negative breast cancer under the two treatments of interest, this may not be the case for other trial settings, e.g., when the new therapy could reduce the extent of the disease even among non-pCR patients, thus improving long-term survival endpoints compared to the control arm.

An imperative 
and practical question   is: how can we use the short term efficacy results on pCR to design a phase III trial with long-term survival endpoints? Based on historical data, we may have reasonable information on the survival distributions among those who achieved pCR, and the survival distributions among those who do not  for similar patient populations. 
Motivated by such challenges, in this paper we aim to design a phase III trial for a  two-sample comparison between the  control group and the intervention group, with respect to a  long-term survival endpoint based on previous information on a short-term binary  endpoint and other available information.

Several authors have addressed how to compare two treatment groups in seamless phase II-III clinical trials with binary and survival endpoints.  
While Inoue et al\cite{Inoue2002}  proposed  a Bayesian approach where the distribution of the time-to-event outcome is specified through a mixture model according to the response, 
Lai et al\cite{Lai2012}  developed group sequential   tests for confirmatory testing based on likelihood ratio statistics for sample proportions and partial likelihood ratio statistics for censored survival data. On the other hand, using an early   response to treatment  as  a potential surrogate endpoint for survival, Chen et al\cite{Chen2020} proposed a joint model for binary  response and a survival time for clustered data,  basing the statistical inference on a multivariate penalized likelihood  method.

The design of clinical trials with survival endpoints taking into account previous information on binary endpoints has received less attention.
Abberbock et al\cite{Abberbock2019} assessed how to relate the improvement in pCR with the effect size in survival in the context of neoadjuvant breast cancer trials.  
In order to  cope with the possibility of a  non-constant hazard ratio, Abberbock et al considered the average of the hazard ratio over the follow-up period  as the effect measure.   
Schoenfeld's formula is  then used to calculate the  required number of events and augmented for the sample size under the exponential distribution.
However, as it has been widely discussed, when the proportional hazards assumption does not hold, the interpretation of the hazard ratio  and  the average hazard ratio might not be straightforward neither  clinically nor statistically.\cite{Kalbfleisch1981,Rauch2018,Royston2011,Royston2013,Zhao2012}  

In this paper, we propose a design for clinical trials with long-term survival endpoints, given  the information on short-term binary endpoints. We distinguish between patients who respond to the binary endpoint, called responders, and those who do not, called non-responders. The anticipated effect size and sample size  are calculated on the basis of the response rate of the binary endpoint, as well as on  the survival  functions for  responders and non-responders in each treatment arm.  The  corresponding survival distribution for each arm is hence a mixture between responders and non-responders, and the common  assumption on the proportionality of the hazard rates is unlikely to be satisfied. To overcome these difficulties we  propose to  consider the restricted mean survival time 
for each treatment arm, and to use their difference as the basis of comparison.  
The difference between the restricted mean survival times has several advantages: it is easily interpretable as the mean difference in the survival times by the end of follow up,\cite{Royston2011,Royston2013} and  it does not require  the proportionality of the hazards.

We present the  effect size and sample size formulae for designing such trials. As shown in this work, the sample size calculation requires previous knowledge on the survival by response distributions and the response rates for each treatment group.   
We additionally outline how the sample size can be determined according to different parameter choices.  
In order to make our proposal easy to use in practice, our method has been implemented in the  \texttt{survmixer} R package, which is available for download on CRAN. \cite{survmixer2021}

This paper is organized as follows. In Section \ref{sect-2settings}, we  introduce the notation and main assumptions. 
In Section \ref{sect-4proposal}, we describe the design of trials using restricted mean survival times and provide the effect size and sample size formulae based on short-term endpoints and survival by response information.
We present the R package \texttt{survmixer} in Section \ref{sect-5implementation} and illustrate our proposal using a neoadjuvant trial  
in Section \ref{sect-6example}. We perform a simulation study in Section \ref{sect-7simulations} to evaluate the performance our proposal.
We conclude the paper with a short discussion.
The R code used in this paper is publicly available on GitHub (\url{https://github.com/MartaBofillRoig/survmixer}).


\section{Notation and Assumptions} \label{sect-2settings}

Consider a randomized clinical trial designed to compare two treatment groups,
control group ($i=0$) and intervention group ($i=1$), 
each composed of $n^{(i)}$ individuals, and denoting by $n=n^{(0)}+n^{(1)}$ the total sample size.  
Suppose that individuals from both groups are followed over the time interval  $[0,T_{end}]$ and are 
compared with respect to a long-term time-to-event outcome  evaluated
within the interval $[0,\tau]$ ($0<\tau \leq T_{end}$), such as event-free survival (EFS).
Let $T_{ij}$ and $C_{ij}$ be the time from randomization to the
long-term  outcome and to censoring, respectively, for each subject
$j=1, ..., n^{(i)}$, $i=0,1$, and let  $S^{(i)}(\cdot)$ be the
survival function of  $T_{ij}$ for  the $i$-th group.
Assume that $T_{ij}$ and $C_{ij}$ are independent.

Assume that we have prior information on a short-term outcome, such as the pCR status.\cite{DeMichele2015,Broglio2016}  
Based on this short-term outcome, each patient who achieved the response is deemed a responder; otherwise, they are considered a non-responder.  
Let $X_{ij}=1$ indicate that the $j$-th patient  in  the $i$-th group   is a responder; otherwise $0$. Let $p^{(i)}$  be  the     probability of having   responded. 

We aim to design a superiority phase III trial for a  two-sample comparison
between the  control group and the intervention group with respect to 
the long-term outcome as we establish  in the following hypothesis:
\begin{eqnarray}\label{H0:S}
	\mathrm{H}_0: S^{(0)}(t) =S^{(1)}(t), \ \ \forall t \in
	[0,\tau] \quad & {\rm vs} &\quad  \mathrm{H}_1: S^{(1)}(t) \geq S^{(0)}(t),
	\ \ \ 
	S^{(1)}(t) > S^{(0)}(t)
	\ \ \  {\rm for \ some \ } t \in
	[0,\tau]. 
\end{eqnarray}  

\subsection{Survival function and effect size in terms of the short-term outcome and  responders}

Let  $S_r^{(i)}(t)=P(T_{ij}>t| X_{ij}=1)$ and $S_{nr}^{(i)}(t) =P(T_{ij}>t| X_{ij}=0)$ denote the  survival functions for responders and non-responders  in the $i$-th group, respectively. The survival function for the long-term outcome in the $i$-th group  can then be expressed as a mixture of them  as follows:  
\begin{eqnarray} \label{mixture-surv}
	S^{(i)}(t)&=&P(T_{ij}>t, X_{ij}=1)+P(T_{ij}>t, X_{ij}=0) 
	=p^{(i)} \cdot S_r^{(i)}(t) +(1- p^{(i)}) \cdot S_{nr}^{(i)}(t). 
\end{eqnarray}
From the survival function given in \eqref{mixture-surv}, the difference in survival functions at $t$ ($t\in[0,\tau]$)  is as follows:
\begin{eqnarray*}
	S^{(1)}(t)-  S^{(0)}(t) 
	&=& p^{(1)} \left ( S_r^{(1)}(t) -S_{nr}^{(1)}(t)\right)- p^{(0)} \left ( S_r^{(0)}(t) -S_{nr}^{(0)}(t)\right) +S_{nr}^{(1)}(t) -S_{nr}^{(0)}(t).
\end{eqnarray*}

Although the hazard ratio  is the most commonly used effect measure  in survival analysis, it relies on the assumption  of  a constant hazard ratio over time between the two groups. 
However, under a mixture model such as \eqref{mixture-surv}, the proportionality of the hazards rarely holds, even if  the survival functions  for  responders and non-responders are exponentially
distributed,\cite{Abberbock2019} as we will illustrate in Section \ref{ES_designcases}.  The expression of the hazard ratio under the mixture model  can be found in the Supplementary Material (Theorem B).

As an alternative to the hazard ratio to quantify the effect of an intervention, we can use the difference of the restricted mean survival times (RMSTs) for each group.  
The RMST is defined as the mean survival time within a specific time window $[0,\tau]$,  and it corresponds to the area under the survival curve until $\tau$: 
\begin{equation} \label{def_rmst}
	K^{(i)}(\tau) = \int_{0}^{\tau} S^{(i)}(t) dt.
\end{equation}
The RMST corresponding to the survival function given in \eqref{mixture-surv} can be expressed as follows:
\begin{eqnarray*} 
	K^{(i)}(\tau)  &=&  
	p^{(i)} K^{(i)}_r(\tau)+ (1-p^{(i)} ) K^{(i)}_{nr}(\tau),  
\end{eqnarray*}
where $K^{(i)}_r(\tau) =\int_{0}^{\tau} S^{(i)}_r(t)   dt $ and
$K^{(i)}_{nr}(\tau) =\int_{0}^{\tau} S^{(i)}_{nr}(t)   dt $ denote the
RMST for the  responders survival and the non-responders survival in the $i$-th treatment arm, respectively. 
The difference between arms in RMSTs is then:
\begin{eqnarray} \label{DifRMST}
	K^{(1)}(\tau) - K^{(0)}(\tau)  &=&  
	\left( p^{(1)} K^{(1)}_r(\tau)-p^{(0)} K^{(0)}_r(\tau)\right)
	+\left( (1-p^{(1)} ) K^{(1)}_{nr}(\tau) -  (1-p^{(0)} ) K^{(0)}_{nr}(\tau) \right),  
\end{eqnarray} 
where   each term in the above sum corresponds to a function of the  responders and for non-responders survival functions. Details for this derivation  are provided in the Supplementary Material (see Theorem A). 

Note that if H$_0$  in \eqref{H0:S} is true, then $K^{(1)}(\tau) -
K^{(0)}(\tau) =0$; whereas, under H$_1$, $K^{(1)}(\tau) - K^{(0)}(\tau)
>0$ and the hypothesis can be rewritten as: 
\begin{eqnarray}\label{H0:K}
	\mathrm{H}_0: K^{(0)}(\tau) =K^{(1)}(\tau)  \quad
	& {\rm vs} &
	\quad  \mathrm{H}_1: K^{(1)}(\tau) > K^{(0)}(\tau).
\end{eqnarray}
From now on and for the rest of the paper, we focus on $\mathrm{H}_0$ and $\mathrm{H}_1$ as stated in \eqref{H0:K}.


\section{Sample Size based on short-term outcomes  and survival-by-responders endpoints} \label{sect-4proposal}   

In this section, we start describing the two-sample test based on the  difference of the RMSTs. We  then provide the expression for  the overall mean survival improvement and discuss different design settings. We end the section with the derivation of the sample size  based on the mixture survival function.

\subsection{Test statistic}

Let  $\hat{S}^{(i)}(\cdot)$ be the Kaplan-Meier estimate of $S^{(i)}(\cdot)$. A consistent estimator of the RMST in \eqref{def_rmst} is given by $\hat{K}^{(i)}(\tau) = \int_{0}^{\tau} \hat{S}^{(i)}(t) dt$. The distribution of $\sqrt{n^{(i)}}\cdot\left( \hat{K}^{(i)}(\tau)-K^{(i)}(\tau)\right)$  is asymptotically  normal  with $0$ mean and limiting variance $\left( \sigma^{(i)}(\tau)\right)^2$, defined as: 
\begin{eqnarray} \label{sigma_rmst}
	\left( \sigma^{(i)}(\tau)\right)^2  &=& -\int_{0}^{\tau}
	\frac{(K^{(i)}(\tau)-K^{(i)}(t))^2}{(S^{(i)}(t))^2\cdot G^{(i)}(t)}
	dS^{(i)}(t) 
	=
	-\int_{0}^{\tau} \frac{(\int_{t}^{\tau} S^{(i)}(u) du)^2}{(S^{(i)}(t))^2\cdot G^{(i)}(t)}  dS^{(i)}(t), 
\end{eqnarray}
where $G^{(i)}(\cdot)$ is the survival function of the censoring
variable  $C_{ij}$ for the $i$-th group.

To test the null hypothesis H$_0$ in \eqref{H0:K}  against the alternative hypothesis H$_1$,   
we consider the statistic: 
\begin{equation}\label{test}
	Z_{s,n} =( {\hat{K}^{(1)}(\tau)-\hat{K}^{(0)}(\tau)})\Big/{\sqrt{\frac{(\hat{\sigma}^{(0)}(\tau))^2}{n^{(0)}}+\frac{(\hat{\sigma}^{(1)}(\tau))^2}{n^{(1)}}}},
\end{equation}
where $\hat{\sigma}^{(i)}(\tau)$ is the estimate of the variance $\left( \sigma^{(i)}(\tau)\right)^2$, which is obtained by substituting  $K^{(i)}(\cdot)$, $S^{(i)}(\cdot)$, and $G^{(i)}(\cdot)$ by its corresponding estimates $\hat{K}^{(i)}(\cdot)$, $\hat{S}^{(i)}(\cdot)$, and $\hat{G}^{(i)}(\cdot)$. 

Let $D(\tau) = K^{(0)}(\tau)-K^{(1)}(\tau)$ be the minimum meaningful effect to be detected. Then, $\hat{D}(\tau)=\hat{K}^{(0)}(\tau)-\hat{K}^{(1)}(\tau)$ is a consistent estimator of $D(\tau)$. 
Moreover, $\sqrt{n}\left((\hat{K}^{(0)}(\tau)-\hat{K}^{(1)}(\tau)) -  (K^{(0)}(\tau)-K^{(1)}(\tau)) \right) $	is asymptotically normal distributed with mean $0$ and variance equal to
$(\sigma^{(0)}(\tau)+\sigma^{(1)}(\tau))^2$. By Slutsky's theorem, the statistic $Z_{s,n}$ is asymptotically $N(0,1)$ under $H_0$ and  asymptotically normal  with mean equal to $D(\tau)\big / \sqrt{((\sigma^{(0)}(\tau))^2+(\sigma^{(1)}(\tau)))^2}$ and unit variance under a fixed alternative equal to $D(\tau)$. 

For further details on the restricted mean survival times, we refer to the work of  Luo et al,\cite{Luo2019} Zhao et al,\cite{Zhao2016}  
Pepe and Fleming,\cite{PepeFleming1989} Gill,\cite{Gill1980} and the references therein.

\subsection{Effect size }\label{EffectSize}

We denote by $\Delta_r(\tau)= K^{(1)}_r(\tau)- K^{(0)}_r(\tau)$ and $\Delta_{nr}(\tau)= K^{(1)}_{nr}(\tau)- K^{(0)}_{nr}(\tau)$ the  mean survival improvement 
of the intervention group over the control group  for responders and non-responders, respectively; 
and by $\Delta_0(\tau)=K^{(0)}_{r}(\tau)- K^{(0)}_{nr}(\tau)$ the mean survival improvement 
of responders against non-responders in the control group. The  treatment effect on the response rate is denoted by $\delta_p=p^{(1)}-p^{(0)}$.  

The overall mean survival improvement  between treatment groups  in \eqref{DifRMST} can be re-expressed as:
\begin{eqnarray} \label{DifRMST2}
	D(\tau) = K^{(1)}(\tau) - K^{(0)}(\tau) &=& p^{(1)}\cdot \Delta_r +  (1-p^{(1)}) \cdot \Delta_{nr} +  (p^{(1)}-p^{(0)})\cdot \Delta_0. 
\end{eqnarray}

Summarizing, the effect size $D(\tau)$  is then  a function of the:  
\begin{itemize}[noitemsep]
	\item $\Delta_r=\Delta_r(\tau)$: Mean survival improvement due to intervention among responders by time $\tau$. 
	\item $\Delta_{nr}= \Delta_{nr}(\tau)$: Mean survival improvement due to intervention among non-responders by time $\tau$. 
	\item $\Delta_0=\Delta_0 (\tau)$: Mean survival improvement of responders versus non-responders in the control group by time $\tau$. 
	\item $\delta_p$: Improvement due to intervention on the response rate. Note that $p^{(1)}=p^{(0)}+\delta_p$.
	\item $p^{(0)}$:  Probability of response in the control group.
\end{itemize}
When designing a future phase III trial, we need to work closely with our medical collaborators to obtain these quantities, which may be procured  from ongoing or finished phase II trials (assessing similar agents) or historical data using prior scientific  knowledge.

\subsubsection{Effect size under different settings} \label{ES_designcases} 

Next, we discuss four different settings
according to whether or not the survival functions between groups are the same for responders and non-responders.
We consider that the response rate of the intervention arm is higher than that of the control arm ($p^{(1)}-p^{(0)} >0$).  

\begin{enumerate}[label=(\Roman*)]
	\item $S_r^{(1)}(t)= S_{r}^{(0)}(t), \ \ S_{nr}^{(1)}(t)=S_{nr}^{(0)}(t), \ \  S_r^{(0)}(t) > S_{nr}^{(0)}(t)$: The survival function for responders is expected to be superior to that of non-responders. Hence, $\Delta_r= K^{(1)}_r(\tau)- K^{(0)}_r(\tau)=0$,
	$\Delta_{nr}= K^{(1)}_{nr}(\tau)- K^{(0)}_{nr}(\tau)=0$, $\Delta_0>0$, and
	\begin{eqnarray*}
		D(\tau) = (p^{(1)} -p^{(0)})\cdot \Delta_0.
	\end{eqnarray*}
	We note that, even in this simple scenario, the two overall survival functions, $S^{(1)}(\cdot)$ and $S^{(0)}(\cdot)$, are unlikely to satisfy the 	proportional hazards assumption. Under the exponential case for both responders and non responders, the overall survival distribution is no longer exponential:
	$$
	S^{(i)}(t) = p^{(i)} \cdot \exp\{-\lambda_{r}^{(i)} \cdot t\} + (1- p^{(i)}) \cdot  \exp\{-\lambda_{nr}^{(i)} \cdot t\},
	$$
	where  $S_r^{(i)}(t)=\exp\{-\lambda_r^{(i)} \cdot t\}$ and $S_{nr}^{(i)}(t)=\exp\{-\lambda_{nr}^{(i)} \cdot t\}$. Moreover, the  hazards ratio between two treatment groups is then:
	\begin{eqnarray*}  
		\mathrm{HR}(t)   
		&=&\frac{p^{(0)} \cdot \exp\{-\lambda_{r}^{(0)} \cdot t\} + (1- p^{(0)}) \cdot  \exp\{-\lambda_{nr}^{(0)} \cdot t\}}{p^{(1)} \cdot \exp\{-\lambda_{r}^{(1)} \cdot t\} + (1- p^{(1)}) \cdot  \exp\{-\lambda_{nr}^{(1)} \cdot t\}}\cdot \frac{p^{(1)} \cdot \lambda_{r}^{(1)}\exp\{-\lambda_{r}^{(1)} \cdot t\} +(1-p^{(1)} ) \cdot
			\lambda_{nr}^{(1)}\exp\{-\lambda_{nr}^{(1)} \cdot t\}}{p^{(0)} \cdot
			\lambda_{r}^{(0)}\exp\{-\lambda_{r}^{(0)} \cdot t\} +(1-p^{(0)} ) \cdot
			\lambda_{nr}^{(0)}\exp\{-\lambda_{nr}^{(0)} \cdot t\}}, 	
	\end{eqnarray*} 
	showing that the hazard rates are not constant over time.
	
	\item $S_r^{(1)}(t)= S_{r}^{(0)}(t), \ \ S_{nr}^{(1)}(t)> S_{nr}^{(0)}(t)$: Survival improvement due to the intervention for  non-responders, but not for responders. 
	This leads to 
	$\Delta_r=0$, $\Delta_{nr}>0$, and hence:
	\begin{eqnarray*}
		D(\tau) = (1-p^{(1)}) \cdot \Delta_{nr} + (p^{(1)}-p^{(0)})\cdot \Delta_0.
	\end{eqnarray*}
	
	\item $S_r^{(1)}(t)> S_{r}^{(0)}(t), \ \ S_{nr}^{(1)}(t)= S_{nr}^{(0)}(t)$: Responders in the intervention  group  have longer survival
	than control group responders, while there is no mean survival improvement 	among non-responders. Thus, we have  $\Delta_r>0$,
	$\Delta_{nr}=0$, and then:
	$$  K^{(1)}(\tau) - K^{(0)}(\tau) = p^{(1)}\cdot \Delta_r + (p^{(1)}-p^{(0)})\cdot \Delta_0. $$
	
	\item $S_r^{(1)}(t)> S_{r}^{(0)}(t), \ \ S_{nr}^{(1)}(t)> S_{nr}^{(0)}(t)$: Both responders  and non-responders of the intervention group have longer survival than those in the control group. Then  $\Delta_r>0$ and $\Delta_{nr}>0$, so that  the overall mean survival improvement between groups is:	
	\begin{eqnarray*}
		D(\tau) =  p^{(1)}\cdot \Delta_r + (1-p^{(1)}) \cdot \Delta_{nr} + (p^{(1)}-p^{(0)})\cdot \Delta_0.
	\end{eqnarray*}
\end{enumerate}

\subsection{Sample size calculation} \label{ss_approach} 

In order to compute the sample size to test \eqref{H0:K} based on the statistic $Z_{s,n}$, given in \eqref{test},  we need information  on the following quantities: i) the responders rates $p^{(0)}$ and $p^{(1)}$  or, alternatively, instead of $p^{(1)}$, the effect given by $\delta_p=p^{(1)}-p^{(0)}$); ii) the responders and non-responders survival functions   $S_r^{(i)}(\cdot)$ and $S_{nr}^{(i)}(\cdot)$, $i=0,1$; and iii)  the survival censoring function, $G^{(i)}(\cdot)$.  

Let $n(p,S_r,S_{nr},G)$ be the sample size needed for running a trial at significance level $\alpha$ with power $1-\beta$.  Note that the  group indicator has been omitted in the notation for short.
The formula for calculating the total sample size  $n(p,S_r,S_{nr},G)$ is given by:  
\begin{eqnarray}\label{SS-mixture}
	n(p,S_r,S_{nr},G) &=&  
	\frac{(z_\alpha+z_\beta)^2}
	{\left( D(\tau)  \right)^2}\cdot \left (\frac{ 		(\sigma^{(0)}(\tau))^2}{\pi}+\frac{(\sigma^{(1)}(\tau))^2}{1-\pi}\right),
\end{eqnarray}
where   $\pi=\lim n^{(0)}/n$, $z_x$ is the $100\times(1-x)$-th percentile of the standard normal distribution,  and  the variance $(\sigma^{(i)}(\tau))^2$ is:
\begin{eqnarray*}
	(\sigma^{(i)}(\tau))^2 &=& 
	-p^{(i)} \int_{0}^{\tau} \frac{(\int_{t}^{\tau} ( S^{(i)}_r(u)  p^{(i)}+S^{(i)}_{nr}(u)(1-p^{(i)} ) ) du)^2}{( S^{(i)}_r(t)  p^{(i)}+S^{(i)}_{nr}(t)(1-p^{(i)} ) )^2\cdot G^{(i)}(t)}  d S^{(i)}_r(t)  \\
	&&
	-(1-p^{(i)} )\int_{0}^{\tau} \frac{(\int_{t}^{\tau} ( S^{(i)}_r(u)  p^{(i)}+S^{(i)}_{nr}(u)(1-p^{(i)} ) ) du)^2}{( S^{(i)}_r(t)  p^{(i)}+S^{(i)}_{nr}(t)(1-p^{(i)} ) )^2\cdot G^{(i)}(t)}  d S^{(i)}_{nr}(t).
\end{eqnarray*}
The derivation of the sample size \eqref{SS-mixture} can be found in the Supplementary Material (see Theorem C).

In order to evaluate $D(\tau)$, we can  employ either \eqref{DifRMST}, where the restricted mean survival times  for responders and non-responders are used, or \eqref{DifRMST2}, where the anticipated survival benefits for responders and non-responders are considered.  However, we notice that, if using \eqref{DifRMST2} for anticipating $D(\tau)$ in $n(p,S_r,S_{nr},G)$,  the anticipation
of  $S_r^{(i)}(\cdot)$ and $S_{nr}^{(i)}(\cdot)$ in
$(\sigma^{(i)}(\tau))^2$ has to be in consonance with the expected
effect sizes  ($\Delta_r,  \Delta_{nr}, \Delta_{0}$).

\subsubsection{Sample size based on interpretable parameters} \label{ssapprox_approach} 

Sample  size  calculations  always  depend  on  a  number  of  factors  that  have  to  be  estimated  from  pilot studies, literature, or educated guesses based on clinical experience. The proposed sample size computation \eqref{SS-mixture} relies on the expected parameters for the responder rates, the survival distributions, and survival censoring function. This prior information might be obtained in terms of different summary statistics. 
In this section, we propose three distinct sets of summaries; choosing one or another would depend on the previous information
on the survival functions for responders and non-responders. These three summary statistics are essentially equivalent under our distributional assumptions.

In this section,  we assume that both responders and non-responders survival functions follow exponential distributions. Similar derivations, if  Weibull distributions are assumed, can be found in the Supplementary Material. The censoring distribution is assumed to be exponential.

Researchers should have the response rate in the control group, $p^{(0)}$, the anticipated effect due to the intervention in the response rate, $\delta_p$, and the scale parameter for the censoring distribution. Furthermore,  one of the following three sets of summaries are needed: 
\begin{itemize}
	\item 
	\textbf{Summary statistics (I): Sample size based on ($m_{r}^{(0)}$, $m_{nr}^{(0)}$, $m_{r}^{(1)}-m_{r}^{(0)}$,  $m_{nr}^{(1)}-m_{nr}^{(0)}$):}\\
	We would need the mean survival time for responders and  non-responders distributions in the control group,
	$m_{r}^{(0)}$ and $m_{nr}^{(0)}$, and the differences in mean survival time
	for responders and non-responders,
	$m_{r}^{(1)}-m_{r}^{(0)}$ and  $m_{nr}^{(1)}-m_{nr}^{(0)}$, respectively.
	
	From there, we could directly translate these anticipated values to the parameters of the exponential distributions and calculate the sample size accordingly using \eqref{SS-mixture}.
	
	\item 
	\textbf{Summary statistics (II): Sample size based on ($S_r^{(1)}(\tau)-S_r^{(0)}(\tau), S_{nr}^{(1)}(\tau)-S_{nr}^{(0)}(\tau)$, $S_r^{(0)}(\tau)$, $S_{nr}^{(0)}(\tau)$):}\\
	We would  require here the $\tau$-year survival rates for responders and non-responders in the control group, $S_r^{(0)}(\tau)$ and $S_{nr}^{(0)}(\tau)$, and 
	the difference in survival functions at $\tau$ for responders and non-responders,  $S_r^{(1)}(\tau)-S_r^{(0)}(\tau)$ and $S_{nr}^{(1)}(\tau)-S_{nr}^{(0)}(\tau)$.

	Based on this information, 
	we could deduce the parameters of the exponential distributions and calculate the sample size according to \ref{SS-mixture}.
	
	\item 
	\textbf{Summary statistics (III): Sample size based on ($S_r^{(0)}(\tau)$, $S_{nr}^{(0)}(\tau)$, $\Delta_r, \Delta_{nr}$):}\\
	We would need the $\tau$-year survival rates for responders and non-responders in the control group, $S_r^{(0)}(\tau)$ and $S_{nr}^{(0)}(\tau)$, and the mean survival improvement for responders and non-responders, $\Delta_r$ and $\Delta_{nr}$. 
	
	Based on this information, we establish  the underlying relationships between the anticipated set of parameters and  the parameters of the exponential distribution 
	and get approximated values for the rate parameters using the Taylor series. 
	Once we have the parameters, we calculate the sample size according to \ref{SS-mixture}.
\end{itemize}

In the Supplementary Material, we stated the formulae that we have used to obtain the distributional parameters from each of the summary statistics.


\section{Implementation} \label{sect-5implementation}

Results in Section \ref{sect-4proposal} allow for the calculation of the sample size and effect size for overall survival based on the response rate and the survival-by-response information.
To make these results accessible to clinical trial practitioners, we have created the R package \texttt{survmixer} (\url{https://github.com/MartaBofillRoig/survmixer}),  which incorporates two main functions:  \texttt{survm\_effectsize} and \texttt{survm\_samplesize} for calculating the effect size (RMST difference) in \eqref{DifRMST2} and the sample size in \eqref{SS-mixture}, respectively. 

In the function \texttt{survm\_effectsize},  the RMST difference  can be computed   based on two different sets of arguments, the choice of which is based on  the parameter \texttt{anticipated\_effects}. 
If \texttt{anticipated\_effects} is \texttt{TRUE},  the overall mean survival improvement is computed based on the formula \eqref{DifRMST2}, and then using the set of arguments ($\Delta_r, \Delta_{nr}$, $\Delta_0$, $p^{(0)}$, $\delta_p$), that is: 
\begin{lstlisting}
survm_effectsize(Delta_r, Delta_0, Delta_nr, delta_p, p0,
anticipated_effects=TRUE)
\end{lstlisting}
where 	\texttt{Delta\_r, Delta\_0, Delta\_nr, delta\_p, p0} are the already introduced parameters $\Delta_r$, $\Delta_{nr}$, $\Delta_0$, $\Delta_p$ and $p^{(0)}$. 

On the other hand, if  \texttt{anticipated\_effects} is \texttt{FALSE}, the overall mean survival improvement is computed according to \eqref{DifRMST} and then 
based on the  set of arguments $(S_r^{(0)}(\cdot), S_r^{(1)}(\cdot), S_{nr}^{(0)}(\cdot), S_{nr}^{(1)}(\cdot), p^{(0)}, \delta_p)$, 
that is: 
\begin{lstlisting}
survm_effectsize(ascale0_r, ascale0_nr, ascale1_r, ascale1_nr, delta_p, p0, 
bshape0=1, bshape1=1, tau, 
anticipated_effects=FALSE)
\end{lstlisting}
where
\texttt{delta\_p, p0} are the effect size and event rate for the response rate ($\delta_p$, $p^{(0)}$);
\texttt{ascale0\_r, ascale0\_nr, ascale1\_r, ascale1\_nr} are the scale parameters for the distribution in both the control  and  intervention groups for responders and non-responders; 
and \texttt{tau} is the end of follow up. 
The responders and non-responders survival functions are assumed to be exponentially distributed. However, they can be assumed to be Weibull distributed by using the arguments \texttt{bshape0,bshape1}, which are the shape parameters in the control  and  intervention groups.

The \texttt{survm\_samplesize} function computes the sample size on the basis of different sets  of summary statistics, as explained in Section \ref{ssapprox_approach}.   
The user can choose which set of summaries to use by means of the argument \texttt{set\_param}. This function can be called for each of the parameter settings by:
\begin{lstlisting}
survm_samplesize(m0_r,m0_nr,diffm_r,diffm_nr,delta_p,p0,tau,ascale_cens,
alpha,beta,set_param=1)
survm_samplesize(S0_r,S0_nr,diffS_r,diffS_nr,delta_p,p0,tau,ascale_cens,
alpha,beta, set_param=2)
survm_samplesize(Delta_r,Delta_nr,S0_r,S0_nr,delta_p,p0,tau,ascale_cens,
alpha,beta, set_param=3)
\end{lstlisting} 
where the arguments:
\begin{itemize}[noitemsep]
	\item \texttt{m0\_r, m0\_nr} are the   mean survival time for responders and non-responders in the control group ($m_r^{(0)}$,$m_{nr}^{(0)}$); 
	\item
	\texttt{diffm\_r, diffm\_nr} are the difference in    mean survival time between group for responders and non-responders ($m_{r}^{(1)}-m_{r}^{(0)}$,  $m_{nr}^{(1)}-m_{nr}^{(0)}$);
	\item
	\texttt{S0\_r, S0\_nr} are the $\tau$-year survival rates for responders and non-responders in the control group ($S_r^{(0)}(\tau)$, $S_{nr}^{(0)}(\tau)$); 
	\item
	\texttt{diffS\_r, diffS\_nr} are the difference in   survival functions at $\tau$ for responders and non-responders ($S_r^{(1)}(\tau)-S_r^{(0)}(\tau), S_{nr}^{(1)}(\tau)-S_{nr}^{(0)}(\tau)$);   
	\item
	\texttt{Delta\_r,  Delta\_nr, delta\_p, p0} are the same arguments that we have in \texttt{survm\_effectsize} ($\Delta_r, \Delta_{nr}$, $\delta_p$, $p^{(0)}$); 
	\item 
	\texttt{ascale\_cens} is the scale parameter for the censoring distribution; 
	\item  \texttt{alpha} and \texttt{beta} are the pre-specified type I and type II errors, respectively. 
\end{itemize}


\section{Motivating Example: The NOAH trial} \label{sect-6example}

In phase III of the NOAH (NeOAdjuvant Herceptin) trial,\cite{Gianni2010,Gianni2014}   
the primary objective was to assess whether neoadjuvant chemotherapy with one year of trastuzumab improved event-free survival as compared with neoadjuvant chemotherapy alone in patients with HER2-positive breast cancer.  
Patients in the NOAH trial were randomly assigned to receive neoadjuvant chemotherapy alone or neoadjuvant chemotherapy plus one year of trastuzumab.  
The primary endpoint  was event-free survival, defined as the time from randomization until disease recurrence, progression, or death from any cause. Secondary endpoints were,  among others, pathological complete response (pCR) in breast tissue  
and overall survival. 

A total of 235 patients with HER2-positive disease were enrolled in the study, of whom
118 received chemotherapy alone  and 117 received chemotherapy plus trastuzumab.
The sample size  was calculated using Shoenfeld's formula  to have $80\%$ power to detect a hazard ratio of $0.545$ on the primary endpoint at a two-sided $\alpha$ level of $0.05$, assuming a median event-free survival of $5.5$ years with trastuzumab plus chemotherapy.  The two treatment groups were compared using the logrank test, and a Cox proportional hazards model was used to estimate the hazard ratios and to test their significance.

For illustrative purposes, we assume  that  a phase III trial is to be  conducted  based on the estimated values of  the  pCR rate in each treatment arm, and $5$-year survival rate 
by pCR status in each treatment arm  observed from the NOAH trial.
The NOAH trial showed that the overall event-free survival  at $5$
years was $58\%$  ($87\%$  for pCR responders and  $38\%$ for
non-responders) in the trastuzumab plus chemotherapy group (trastuzumab group for short), and it was $43\%$ in the
neoadjuvant chemotherapy group ($55\%$ for pCR responders and $41\%$
for non-responders); whereas the pCR rate was $0.38$ in the trastuzumab group and $ 0.19$ in the chemotherapy group.
Based on  these results  as model inputs  
and assuming  exponential distributions for
the survival  and censoring distributions, the  mean event-free
survival value is  $8.36$ and $5.61$ years for
responders and non-responders in the chemotherapy group, respectively, and 
$35.90$ and $5.17$ years for responders and non-responders in
the trastuzumab group, respectively. 

Observe that non-responders in the chemotherapy group have slightly larger survival than the non-responders in the  trastuzumab group.  For the purpose of clarity, we assume for the illustration that both responders and non-responders follow the same survival function with the mean equals to $5.17$ years.
We additionally assume  equal exponential censoring
distributions for the two groups with the mean equals to $7$ years.

\begin{figure}[h]
	\centering
	\includegraphics[width=0.8\linewidth]{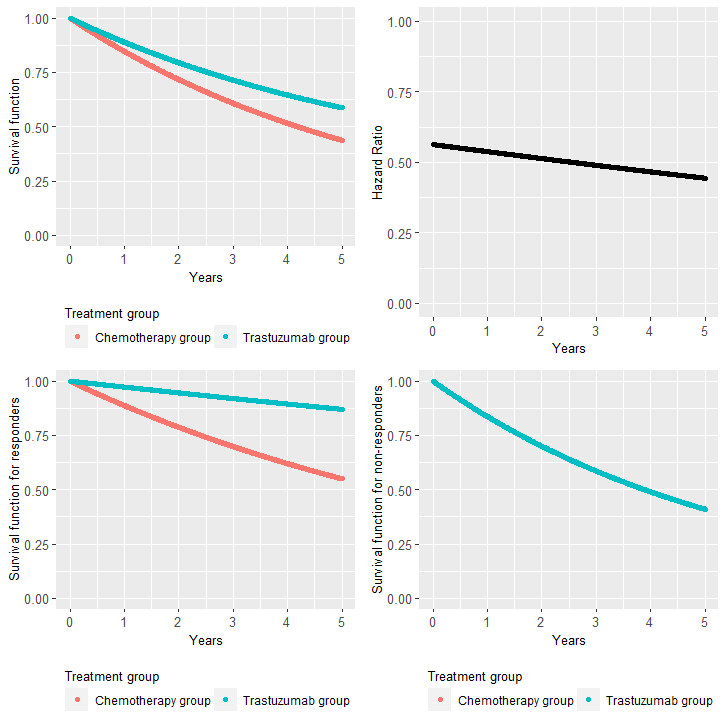}
	\caption{
		The top-left plot depicts the survival functions for event-free survival   for each treatment arm ($S^{(0)}(\cdot)$  and $S^{(1)}(\cdot)$), calculated by means of \eqref{mixture-surv} and using the values in Table \ref{table_results}.  The top-right plot is the hazard ratio over the follow-up period (calculated using formula (1.c) in the Supplementary Material). The bottom-left and bottom-right plots are the survival functions by response in each treatment arm  ($S_r^{(0)}(\cdot)$  and $S_r^{(1)}(\cdot)$ for responders, and  $S_{nr}^{(0)}(\cdot)$  and $S_{nr}^{(1)}(\cdot)$ for non-responders). 
	}
	\label{fig:dsitributionsdesign}
\end{figure}

Table \ref{table_results} provides the values we are using to compute the sample size in this hypothetical  new study. 
Figure \ref{fig:dsitributionsdesign} shows the survival functions by response for each treatment arm, as well as the survival functions for the whole population calculated as the mixture of responders and non-responders given in \eqref{mixture-surv}.  
Figure \ref{fig:dsitributionsdesign}  also plots the hazard ratio between treatment arms over time  (given in formula (1.c) in the Supplementary Material); observe that it  lies between $0.45$ to $0.65$.  This departure from constancy is the consequence of the different survival patterns of pCR responders and non-responders within the  survival mixture model.

\begin{table}[h]
	\centering
	\caption{ 
		\textbf{Anticipated parameters for the design of the phase III trial:}
		Probability of the binary endpont pCR in each treatment group ($p^{(0)}$ and $p^{(1)}$); $5$-year survival rates for responders and non-responders in each treatment group, $S_r^{(i)}(\tau)$ and $S_{nr}^{(i)}(\tau)$, $i=0,1$; and the mean survival time for the exponential functions by treatment arm, by pCR response, and for the censoring distribution.} 
	\label{table_results}
	\begin{tabular}{clc}
		\toprule
		\textbf{Parameters} &  & \textbf{Anticipated values} \\ 
		\hline
		Probability of achieving pCR & in  chemotherapy group & $0.19$\\
		& in  trastuzumab group & $0.38$ \\
		\hline
		$5$-year survival rate &  for responders in  trastuzumab group & $0.87$ \\
		&  for responders in  chemotherapy group & $0.38$ \\
		&  for non-responders in  trastuzumab group & $0.41$ \\ 
		&  for non-responders in  chemotherapy group & $0.41$ \\  
		\hline
		Mean survival time & for responders in  trastuzumab group & $35.90$ \\
		& for responders in  chemotherapy group & $8.36$ \\
		& for non-responders in  trastuzumab group & $5.61$ \\ 
		& for non-responders in  chemotherapy group & $5.61$ \\
		& for censoring & $7$ \\ 
		\bottomrule  
	\end{tabular}
\end{table} 

We guide the reader on how to get the expected overall mean survival improvement and how to size the phase III trial using the \texttt{survmixer} package (in Section \ref{sect-5implementation}).
We calculate the overall mean survival improvement (RMST difference) between groups by means of the function  \texttt{survw\_effectsize}. To do so, we consider the values of the mean survival time for responders and non-responders in each group, the probability of achieving pCR response in the control group, and the response difference between groups (see Table \ref{table_results}). As shown below, the function \texttt{survw\_effectsize}  returns the overall mean survival improvement and the mean survival improvement that
would be assumed for both responders and non-responders.
\begin{lstlisting}
survm_effectsize(ascale0_r=8.37,ascale0_nr=5.61,ascale1_r=35.90,ascale1_nr=5.61,
delta_p=0.19,p0=0.19,tau=5)

##                        Parameter     Value
## 1                RMST difference     0.43 
## 2     RMST difference responders     0.90 
## 3 RMST difference non-responders     0.00 
## 4            Response difference     0.19
\end{lstlisting} 
%
The resulting overall mean survival  improvement of trastuzumab over neoadjuvant chemotherapy is $0.43$ years. This difference is mainly because of the mean survival improvement among responders, since the   mean survival improvement between groups ($\Delta_r$) in the patients that responded
is $0.90$ years, whereas there is no improvement for non-responders, $\Delta_{nr}=0$.

Using the difference in survival functions between groups at $\tau=5$ for both responders and non-responders, the probability of response in the control group, and the improvement on the response rate, we calculate the sample size to have $80\%$ power to detect a difference in event-free survival  between the two treatment arms over $5$ years of follow up. We employ the \texttt{survw\_samplesize} function to compute the sample size according to \eqref{SS-mixture}, obtaining that a total sample size of $466$ is needed.
\begin{lstlisting}
survm_samplesize(S0_r=0.55,S0_nr=0.41,diffS_r=0.32,diffS_nr=0,p0=0.19,delta_p=0.19,
ascale_cens=7,tau=5,alpha=0.05,beta=0.2,set_param=2)

##         Parameter       Value
## 1     Sample size       465.98
## 2 RMST difference       0.43
\end{lstlisting}
%

We performed a simulation study to compare the statistical power of the NOAH trial and the one we proposed using the parameter inputs derived from the NOAH trial.  
We assumed exponential distributions for the survival functions for each of the four subgroups, considered the parameters  in Table \ref{table_results}, and replicated $10000$ times to estimate the power to detect a statistically significant  difference in event-free survival  between the two arms. 
When simulating trials  of size $n=235$, as in the NOAH trial, we obtained an empirical power of $0.33$ using the log-rank test and an empirical power of $0.41$ using the RMST test (in \eqref{test}) to detect an improvement in event-free survival  of trastuzumab plus chemotherapy over chemotherapy alone.
On the other hand, when using trials of size $n=466$, we obtained an empirical power of $0.76$ using the log-rank test and an empirical power of $0.80$ using the RMST test.
Note that in both situations our approach leads to higher powers. 
We have additionally evaluated the empirical power under various censoring percentages. The results (not included) show that the empirical powers using the RMST test were, all of them, around 0.80 and higher than the ones using the log-rank test.


\section{Simulation Study} \label{sect-7simulations} 

\subsection{Design }

In this section, we conduct additional simulation studies to evaluate the performance of the proposed sample size calculation in terms of the significance level and the power. 
We simulated a short-term binary endpoint according to the probability of responding to the treatment in control arm $p^{(0)}$ and a difference in response rate between arms of $\delta_p=p^{(1)}-p^{(0)}$.  
For the time-to-event endpoint, we generated the survival times from Weibull distributions for the responders and non-responders survival distributions with scale parameters $a_r^{(i)}$ and $a_{nr}^{(i)}$, respectively, and common shape parameter $b$; that is:  
\begin{eqnarray} \label{sim:distributions}
	S_r^{(i)} = \exp\{-(t/a_r^{(i)})^{b} \}
	\hspace{5mm} \text{and} \hspace{5mm} 
	S_{nr}^{(i)} = \exp\{-(t/a_{nr}^{(i)})^{b}\}.
\end{eqnarray}
The censoring distributions were assumed equal between groups and exponential with scale parameter $a_{cens} =  2 \cdot m_{nr}^{(0)}$, where $m_{nr}^{(0)}$ is the mean of the non-responders in group $0$, that is, $m_{nr}^{(i)} = a_{nr}^{(i)}\cdot\Gamma(1+1/b)$. The same assumption was made in Hatzis et al.\cite{Hatzis2016}

The parameter values used for the simulations are found in Table \ref{table_sim_scenarios}.  We have only considered scenarios that produce  realistic situations and, in particular, that satisfy that $\Delta_r\geq0$, $\Delta_{nr}\geq0$, and $\Delta_{0}\geq0$.
The set of scenarios we considered is available in the GitHub repository \href{https://github.com/MartaBofillRoig/survmixer}{\texttt{survmixer}}.  
For each one of these scenarios, we computed the required sample size
using \eqref{SS-mixture} 
for a one-sided test with
power $1-\beta=0.80$  
at  significance level $\alpha=0.05$. 
Only those scenarios that result in sample sizes between $100$ and $5000$ were taken into account. 
The total number of scenarios considered was $144$.

We performed  $1000$ replications for each configuration and evaluated the power and the significance level by  using  the RMST test in \eqref{test}. 
For comparison purposes,  we also present the results using the log-rank test.

\begin{table}[h!]
	\caption{Simulation scenarios. When simulating under the null hypothesis, we have considered that $a_{nr}^{(1)}=a_{nr}^{(0)}$, $a_r^{(1)}=a_r^{(0)}$, and $\delta_p=0$. When simulating under the alternative hypothesis, we have not considered all possible combinations, we restricted our calculations to those scenarios that yield $\Delta_r\geq0$, $\Delta_{nr}\geq0$, and $\Delta_{0}\geq0$, and whose resulting sample sizes are between $100$ and $5000$. }  
	\centering
	\begin{tabular}{ccc}
		\toprule 
		& \textbf{Parameters} & \textbf{Values}  \\ 
		\bottomrule 
		& $\tau$ & $10$ \\ 
		& $p^{(0)}$ & $0.1,0.3$ \\ 
		& $b$ & $1,2$ \\  
		& $a_r^{(0)}$ & $15, 18, 20,22, 50$ \\  
		& $a_{nr}^{(0)}$ & $5,6,7,8,9,10,11, 15,17,29$ \\  \bottomrule
		Under the null hypothesis & $a_{nr}^{(1)}=a_{nr}^{(0)}$ &  \\ 
		& $a_r^{(1)}=a_r^{(0)}$ &  \\   
		& $\delta_p$ & $0$ \\ 
		\bottomrule
		Under the alternative hypothesis &  $a_{nr}^{(1)}$ & $5,7,8,9,10,11,13,15,17,19,20,29,35,40$ \\ 
		& $a_r^{(1)}$ & $22, 25,  50, 55,62, 65$ \\   
		& $\delta_p$ & $0.1,0.3$ \\  \bottomrule
		& $\alpha$ & $0.05$ \\
		& $\beta$ & $0.2$ \\
		\bottomrule 
	\end{tabular}
	\label{table_sim_scenarios} 
\end{table}

\subsection{Results} 

The results yield to RMST differences for the overall survival  between $0.15$ and $1.82$, with median $0.60$; and to sample sizes between $125$ and $4824$, with median $685$. 
Figure 1 in the Supplementary Material summarizes the sample sizes and
effect sizes obtained for the overall survival under the considered
scenarios with respect to the settings
I to IV discussed in Section \ref{ES_designcases}.
We notice that the scenarios corresponding to settings I and III  are the ones with smaller effect sizes, thus requiring larger sample sizes to achieve the same power.

\begin{figure}[h]
	\centering
	\includegraphics[width=0.6\linewidth]{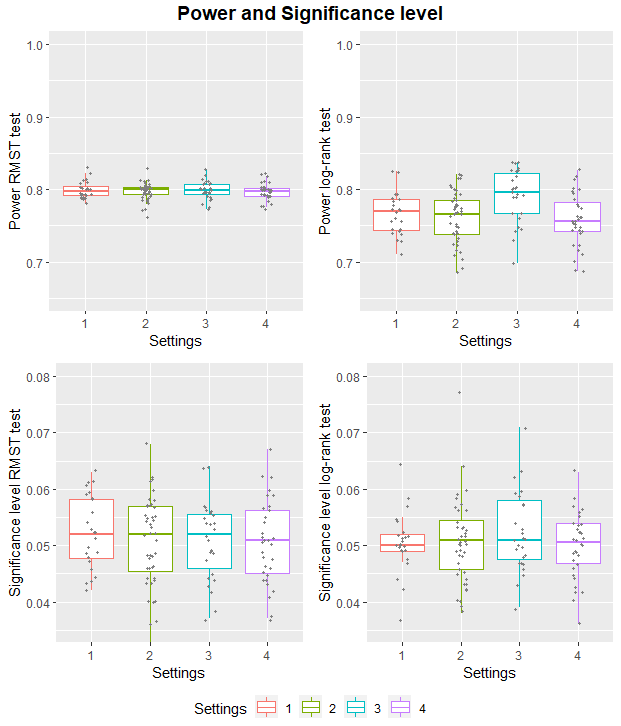}
	\caption{Empirical statistical power and empirical significance level using the RMST test in \eqref{test} and using the log-rank test
		according to the settings  discussed in Section \ref{ES_designcases}.} 
	\label{fig:resultsph}
\end{figure}

We obtained empirical powers with median  $0.80$ (standard deviation equals $0.012$)   using the RMST test and with median $0.77$ ($0.037$)   when using the log-rank test; empirical sizes with median $0.051$ (standard deviations equal $0.007$) when using both the RMST test and the log-rank test.  
Figure \ref{fig:resultsph} shows boxplots of the empirical power and the significance level when using the RMST test and the log-rank test using the same sample sizes and according to the settings in Section \ref{ES_designcases}. We notice that the power obtained using the RMST test is centered around $0.80$, and it has a small variability. 
When comparing with the results using the log-rank test, we observe that the power using the log-rank test is in general less than $0.80$, and there is a greater variability as compared with the results using the RMST test.
The empirical significance level is close to the type I error $0.05$ using both tests.

The simulations presented here illustrate the sample size performance in balanced placebo-controlled trials (1:1 ratio trials). Furthermore, we have studied the sample size for unbalanced designs. In the Supplementary Material, we present the results when the number of patients assigned to the treatment group is higher than in  the  control  group.  We  observe  that  there  is  no  difference  in  the  sample  size  properties  in  terms  of  the significance level and power in unbalanced designs as compared to the ones shown in this section for balanced designs.

\subsection{Sensitivity analysis with respect to model assumptions} 

We  have  evaluated  the  robustness  of  our  design  to  deviations  from  the  model  assumptions.  For  this purpose, we have simulated scenarios under the Weibull distribution and assessed the empirical  $\alpha$ and power obtained if the study would have been designed assuming the exponential distribution. This set of simulations uses the scenarios in Table \ref{table_sim_scenarios}, simulates the data as  Weibull with shape parameters $b=1.5$ and $2$, and designs the study as exponential  ($b=1$). 

The results show that the significance is maintained despite incorrect model specification. On the other hand, the power obtained is, in general, greater than $0.80$.  Specifically, we get: empirical sizes with median $0.051$ and $0.051$  when using the RMST test with $b=1.5$ and $2$, respectively, and $0.052$ and $0.050$ when using the log-rank test; we get empirical powers with median $0.90$ and  $0.90$ using the RMST test with $b=1.5$ and $2$, respectively, and  $0.89$ and $0.88$ using the log-rank test.

\section{Discussion}

Pathologic complete response (pCR) is a common primary endpoint for a phase II trial or for accelerated approval of neoadjuvant cancer therapy. 
If granted,  the regulatory agencies FDA\cite{FDA2014} and EMA\cite{EMA2014} require a two-arm confirmatory trial  to demonstrate the efficacy with a long-term survival outcome, such as  overall survival.   
Because there is not direct way to relate the effects on pCR with the effects on the survival endpoint,   sample size calculation relying on anticipated values for OS is not straightforward.   
In this work, we have approached how to size trials for a survival endpoint based on previous information on a short-term binary endpoint such as pCR. Our proposal is built on the mixture distribution between the binary response and the survival by response and uses the difference of restricted mean survival times (RMSTs) as the effect measure to compare the two treatment arms. We show that both the sample size and the effect size can be written in terms of the probability of response of each treatment arm, and the responders and non-responders survival functions.  The proposed design is suitable for the designing  of long-term phase III neoadjuvant trials given short-term binary outcomes, since, among other reasons, it does not rely on the proportional hazard assumption.

We provide methods along with the  corresponding software to  design those  trials that choose to use the RMST for long-term survival when the proportionality of the  hazards  does  not  hold.   The difference between RMSTs as an alternative summary measure  to the hazard ratio is especially appropriate when the proportional hazards assumption is in doubt. \cite{Royston2011,Royston2013} 
The computation of the RMSTs and their variances involves numerical integrals whose analytic solutions are usually hard to obtain and makes the sample size calculation difficult.  
We have proposed   sample size calculations for planning the trial using RMSTs    based on interpretable parameters without resorting to simulation. These calculations provide an explicit solution to compute the sample size   under the assumption that  the survival function of both responders and non-responders follows a  commonly used parametric distribution such as exponential or Weibull. All the necessary sample size calculations have been implemented in the R package \texttt{survmixer}. 
The proposed sample size formulae are  based on asymptotic results for the RMST difference, and  are intended to be used in phase III trials where the sample size is expected to be modest or large.

The RMST depends on the choice of the time window, which should be based on clinical considerations and specified in the study protocol. The choice might be particularly crucial in trials with a small number of patients at the end of follow-up (or truncation time). The protocol must also incorporate the rationale to employ if the last observed time would occur before the truncation timepoint.\cite{Horiguchi2020}


\subsection*{Acknowledgements} 

We would like to thank the referees for their valuable comments and suggestions.
This work was supported by the Ministerio de Econom\'ia y Competitividad (Spain) under Grants PID2019-104830RB-I00 and MTM2015-64465-C2-1-R (MINECO/FEDER); the Departament d'Empresa i Coneixement de la Generalitat de Catalunya (Spain)  under Grant 2017 SGR 622 (GRBIO); and the Ministerio de Econom\'{i}a y Competitividad (Spain), through the Mar\'{i}a de Maeztu Programme for Units of Excellence in R\&D  under Grant MDM-2014-0445 to M. Bofill Roig.
Y. Shen is partially supported by Biostatistics Shared Resource through Cancer Center Support Grant (CA016672), from the National Cancer Institute, National Institutes of Health. 


\end{document}